\documentclass[prd,floatfix,superscriptaddress,12pt,tightenlines
]{revtex4-1}
\usepackage[english]{babel}
\usepackage{amsmath,amssymb,amsbsy,amstext,amsthm,booktabs,simplewick,exscale,relsize,graphicx,amsfonts,upgreek}
\usepackage{multirow,dcolumn,bm,enumerate,url,hyperref,wasysym}
\usepackage{color,soul}

\newcommand{\beq}{\begin{equation}}
\newcommand{\eeq}{\end{equation}}
\newcommand{\ba}{\begin{array}}
\newcommand{\ea}{\end{array}}
\newcommand{\bea}{\begin{eqnarray}}
\newcommand{\eea}{\end{eqnarray} }
\newcommand{\nn}{\nonumber}
\usepackage[utf8]{inputenc}

\begin{document}

\title{Light Scalars and Dark Photons \\ in Borexino and LSND Experiments}

\author{Maxim Pospelov}\email{mpospelov@perimeterinstitute.ca}
\affiliation{Department of Physics and Astronomy, University of Victoria, Victoria, BC V8P 5C2, Canada}
\affiliation{Perimeter Institute for Theoretical Physics, Waterloo, ON N2J 2W9, Canada}

\author{Yu-Dai Tsai}\email{yt444@cornell.edu}\email{ytsai@fnal.gov}
\affiliation{Fermilab, P.O. Box 500, Batavia, IL 60510, USA \\
Fermi National Accelerator Laboratory, Batavia, IL 60510, USA}

\begin{abstract} 
Bringing an external radioactive source close to a large underground detector can significantly advance sensitivity not only to sterile neutrinos but also to ``dark'' gauge bosons and scalars. Here we address in detail the sensitivity reach of the Borexino-SOX configuration, which will see a powerful (a few PBq) $^{144}$Ce$-^{144}$Pr source installed next to the Borexino detector, to light scalar particles coupled to the SM fermions. The mass reach of this configuration is limited by the energy release in the radioactive $\gamma$-cascade, which in this particular case is 2.2 MeV. Within that reach one year of operations will achieve an unprecedented sensitivity to coupling constants of such scalars, reaching down to $g\sim 10^{-7}$ levels and probing significant parts of parameter space not excluded by either beam dump constraints or astrophysical bounds. Should the current proton charge radius discrepancy be caused by the exchange of a MeV-mass scalar, then the simplest models will be decisively probed in this setup. We also update the beam dump constraints on light scalars and vectors, and in particular rule out dark photons with masses below 1 MeV, and kinetic mixing couplings $\epsilon \gtrsim 10^{-5}$. 
\end{abstract}

\maketitle




\section{Introduction}
\label{sec:intro}

Search for light weakly coupled states undergoes a revival in recent years \cite{Hewett:2012ns}. There has been increased interest in models that operate with light sterile neutrinos, axion-like particles, dark photons, and dark scalars that can be searched for in a variety of particle physics experiments. For a representative but incomplete set of theoretical ideas see, {\em e.g.}
\cite{Silveira:1985rk,Holdom:1985ag,Boehm:2003hm,Boyarsky:2009ix,Pospelov:2007mp,Gninenko:2010pr,Arvanitaki:2009fg,ArkaniHamed:2008qn,Jaeckel:2010ni}. With more emphasis placed on the intensity frontier in recent years, experimental searches of exotic light particles are poised to continue \cite{Alexander:2016aln}. 

Some of this interest is cosmology-driven, exploiting possible connection of light particles to dark matter, or perhaps to a force that mediates 
interactions between Standard Model and dark matter particles \cite{Boehm:2003hm,Pospelov:2007mp,ArkaniHamed:2008qn}. 
In many cases, the interest in light new states is motivated by ``anomalous'' results from previous experiments. The representative anomalies in that respect are the discrepancy in muon $g-2$ measurements \cite{Bennett:2006fi}, puzzling outcomes of some short baseline neutrino oscillation experiments \cite{Aguilar:2001ty,Aguilar-Arevalo:2013pmq,Conrad:2016sve},
and most recently the discrepancy of the charge radius of the proton measured with the muonic and electronic probes \cite{Pohl:2010zza,Pohl:2013yb}. 

One of the most promising avenues for exploring very light and very weakly coupled states is 
by performing experiments in either deep underground laboratories, where the external backgrounds are 
very low, or large detectors usually built for the purpose of studying solar neutrinos. 
With the solar neutrino program currently measuring the last components of the neutrino flux, the usage of these large detectors shifts onto new applications. Thus, the KAMLAND and SNO+ detectors are (or will be) 
used to study double beta decays of Xe and Te isotopes \cite{KamLAND-Zen:2016pfg,Andringa:2015tza}. The Borexino detector will see the expansion of its program to include the sterile neutrino searches 
when new powerful external beta-decay sources are placed nearby \cite{Borexino:2013xxa}. 

There are also interesting proposals based on a possible usage of accelerators underground. Currently, relatively modest accelerators in terms of the energy and current intensity are used in the underground laboratories for measuring the nuclear-astrophysics-relevant reactions \cite{Gustavino:2016txw} or for calibration purposes \cite{Nakahata:1998pz}. These efforts can be significantly expanded.
Powerful accelerators next to large neutrino detectors could open a new way of exploring the nature of light weakly coupled sectors \cite{Aberle:2013ssa,Abs:2015tbh,Izaguirre:2014cza,Kahn:2014sra,Izaguirre:2015pva}.

In this note, we concentrate on the Borexino-SOX project that uses a radioactive $^{144}$Ce$-^{144}$Pr source in close proximity to the detector. The source produces a large number of electron antineutrinos, 
and their signals inside the Borexino as a function of the distance from the source can reveal or constrain 
sterile neutrinos with commensurate oscillation length. In addition, it has already been pointed out that 
the same configuration will be sensitive to the emission of light scalar (or vector) particles in the transitions 
between the nuclear levels in the final point of the $\beta$-decay chain \cite{Izaguirre:2014cza}. 

This note revisits the question of sensitivity of Borexino-SOX to light particles, including dark scalars and newly considered below-MeV dark photons, and updates several aspects of \cite{Izaguirre:2014cza}. We significantly expand the sensitivity reach by taking into account the decays of light particles inside the Borexino detector. Only scalar scattering on electrons was taken into account in the previous consideration. 
In addition, we update the current leading bounds on dark scalars and dark photons by considering the LSND measurements of the elastic electron-neutrino cross section 
\cite{Auerbach:2001wg, Aguilar:2001ty}. Dark photons with masses below 1 MeV can be ruled out with kinetic mixing coupling $\epsilon \gtrsim 10^{-5}$

To have a more specific target in terms of the light particles, in section \ref{sec:model}, we introduce a light scalar coupled to leptons and protons,
which might be responsible for the resolution of the $r_p$ discrepancy \cite{TuckerSmith:2010ra}.
In section \ref{sec:B-SOX}, we calculate the production rate of the 
scalars by relating it to the corresponding nuclear transition rate of $^{144}$Nd. Taking into account the decay and the Compton absorption of the scalars inside the detector we arrive at the expected counting rate, and derive the sensitivity to coupling constants within the 
mass reach of this setup. 
Existing constraints on such light scalars are considered in section \ref{sec:cons}. 
In section \ref{sec:dph}, we study the sensitivity reach of the Borexino-SOX setup in probing light dark photons between a few hundred keV to 1 MeV with a small kinetic mixing (in full awareness of the fact that such light dark photons are disfavored by cosmology). 
We reach the conclusions in section \ref{sec:conclusion}. 

\section{Simplified model of a light scalar and the proton size anomaly}
\label{sec:model}

Following the rebirth of interest in dark photons, other models of light bosons have been closely investigated. In particular, scalar particles are quite interesting, not least because 
they are expected to couple differently to particles with different masses. 
While it is difficult to create a simple and elegant model of dark scalars with MeV range masses, some attempts have been made in refs. \cite{Chen:2015vqy,Batell:2016ove}. We will consider a simplified Lagrangian at low energy in the following form, 
\begin{align}\label{lag_rp}
\mathcal{L}_\phi=\frac{1}{2}(\partial_\mu \phi)^2-\frac{1}{2}m^2_\phi \phi^2 + (g_p \bar{p}p+g_n \bar{n}n+g_e \bar{e}e+g_\mu \bar{\mu}\mu+g_\tau \bar{\tau}\tau)\phi.
\end{align}
In principle, such Lagrangian can be UV-completed in a variety of ways, although it is difficult to maintain both sizable couplings and 
small scalar mass $m_\phi$. In this study, we will not analyze constraints related to UV completion, concentrating instead 
only on the low-energy physics induced by (\ref{lag_rp}).
This simplified Lagrangian with MeV/sub-MeV scalars was proposed in Ref. \cite{TuckerSmith:2010ra} 
(see also \cite{Barger:2010aj,Batell:2011qq,Karshenboim:2014tka,Liu:2016qwd}) 
to explain a 7$\sigma$ disagreement between the measurements of the proton-charge radius using $e - p$ systems 
and the more precise  muonic Hydrogen Lamb shift determination of $r_p$. 
					
More recent data with the Lamb shift in muonic deuterium \cite{Pohl1:2016xoo} show no additional significant 
deviations associated with the neutron, so that the new physics 
interpretation of the anomaly prefers $g_n/g_p \ll 1$. Therefore, we will limit our considerations to $g_n = 0$ case, which will also 
remove all constraints associated with neutron-nucleus scattering \cite{Barbieri:1975xy}. 
Of course, the real origin of the $r_p$  discrepancy is a hotly debated subject, and new physics is perhaps a solution of ``last resort''. 

Introducing the product of couplings,  $\epsilon^2 \equiv g_e g_p/e^2$, one can easily
calculate corrections to the energy levels of muonic atoms due to the scalar exchange.  When interpreted as an effective correction to 
the extracted proton radius from the hydrogen and muonic hydrogen,  this scalar exchange gives
\begin{align}\label{radius_correction}
\Delta r^2_p|_{e\rm H}=-\frac{6\epsilon^2}{m^2_\phi},\; \Delta r^2_p|_{\mu \rm H}=-\frac{6\epsilon^2(g_\mu/g_e)}{m^2_\phi}f(am_\phi)
\end{align}
$f=x^4(1+x)^{-4}$
and $a \equiv (\alpha m_\mu m_p)^{-1} (m_\mu + m_p)$ is the $\rm \mu H$ Bohr radius. For the modifications of the deuterium energy levels, 
one should make $m_p \to m_{\rm D}$ substitution. 

The observed difference \cite{Pohl:2010zza} is
\begin{align}\label{radius_diff}
\Delta r^2_p|_{\rm eH} - \Delta r^2_p|_{\rm \mu H} = -\: 0.063 \pm 0.009 \;\rm fm^2,
\end{align}
and can be ascribed to new physics, provided that it breaks lepton universality.
In particular, it may originate from the $g_\mu  \gg  g_e$ hierarchy, which would be expected from a scalar model. 
For simplicity, we will assume the  mass-proportional coupling constants to the leptons and proton, 
thus $g_e = (m_e/m_\mu)g_\mu$, $g_\tau = (m_\tau/m_\mu)g_\mu$, $g_p = (m_p/m_\mu)g_\mu$, and 
plot the preferred parameter curve in Fig. \ref{fig:sen_con} in green color on the $\epsilon^2 - m_\phi$ plane.

The best part of the new physics hypothesis is that it is ultimately testable with other experimental tools, of which there are many. The most direct way of discovering or limiting such particles is their productions in subatomic experiments with subsequent detection of new particle scattering or decay. 
The MeV-range masses suggested by the $r_p$ anomaly make nuclear physics tools preferable. Such light scalars can be produced in nuclear transitions, and in the next section, we calculate their production in the gamma decay of selected isotopes that are going to be used in the search for sterile neutrinos. 

\section{Borexino-SOX experiment as a probe of scalar sector}
\label{sec:B-SOX}

Here we consider the Borexino-SOX setup in which a radioactive $^{144}$Ce$-^{144}$Pr source will be placed 8.25 meters away from the center of the Borexino detector.

The decay of $^{144}$Ce goes through $^{144} {\rm Ce} \rightarrow \beta \nu^{-} +^{144}$Pr and then $^{144}$Pr$\rightarrow \beta \nu^{-} + ^{144}{\rm Nd}(^{144}{\rm Nd}^{*})$. 
A fraction of the decays results in the excited states of $^{144}{\rm Nd}^{*}$ that $\gamma$-decay to the ground state.  
Then a small fraction of such decays will occur via an emission of a light scalar, 
\begin{equation}
^{144}{\rm Nd}^{*} ~\rightarrow ~^{144}{\rm Nd} + \phi.
\end{equation}

Small couplings of $\phi$ make it transparent to shielding and long-lived relative to the linear scale of the experiment. 
Nevertheless, very rare events caused by the scalar can still be detected by the Borexino detector. 
The main processes via which such scalar can deposit its energy are: 
\begin{align}
\label{three_processes}
e \phi \rightarrow e \gamma&, {\rm~Compton~absorption}\nonumber
\\ 
\phi \rightarrow \gamma \gamma\; &, {\rm~ diphoton~decay}\nonumber 
\\ 
\phi \rightarrow e^{+} e^{-}&, {\rm~ electron-positron~decay}
\end{align}
In what follows we put together an expected strength of such signal, starting from the probability of the scalar emission. 

\subsection{Emission of scalars in nuclear transitions}

Let us find the probability of scalar particle emission in radioactive decays as a function of its mass and coupling.
About one percent of the $^{144}$Ce $\beta$-decays to the 2.185 MeV metastable state of Nd.
This excited state, Nd$^{*}$, then transitions to lower energy states via 1.485 MeV and 2.185 MeV gammas with approximately $30 \%$ and $70\%$ branching ratios \cite{Nakahata:1998pz}. 

{\em Ab initio} calculation of a nuclear decay with an exotic particle in the final state could be a nontrivial task. 
Here, we benefit from the fact that the transition of interest ($^{144}{\rm Nd}^{*} \rightarrow ^{144}{\rm Nd}$) are E1 and the scalar coupling to neutrons is zero, which allows us to link the emission of the scalar to that of the $\gamma$-quanta and thus bypass complicated nuclear physics. 

In the multipole expansion, the relevant part of the interaction Hamiltonian with photons is almost the same form as the corresponding 
counterpart of the scalar interaction, 
\begin{equation}
H_{{\rm int},\gamma}\simeq e \omega A_0\sum_p(\vec{\epsilon}\vec{ r_p});~~
H_{{\rm int},\phi}\simeq g_p \sqrt{\omega^2-m_\phi^2} \phi_0\sum_p(\vec{n}\vec{ r_p}),
\end{equation} 
where $A_0,~\phi_0$ are the amplitudes of the outgoing photon and scalar waves, $\vec{\epsilon}$ and $\vec{n}$ are the unit vectors of 
photon polarization and the direction of the outgoing waves, and the sum is taken over the protons inside the nuclei. After squaring the amplitudes 
induced by these Hamiltonians, summing over polarizations and averaging over $\vec{n}$, we arrive at both rates being proportional to the 
the same square of the nuclear matrix element, $\langle \sum_p \vec{ r_p} \rangle$. In the ratio of transition rates it cancels, leaving us with 
the desired relation
\begin{equation}
\frac{\Gamma_{\phi }}{\Gamma_{\gamma,~E1} } = \frac{1}{2} \left( \frac{g_p}{e}\right)^2\left(1-\frac{m_\phi^2}{\omega^2}\right)^{3/2}.
\end{equation}
All factors in this rate are very intuitive: besides the obvious ratio of couplings, 
the 1/2 factor reflects the ratio of independent polarizations for a photon and a scalar, while $(1-m_{\phi}^2/\omega^2)^{3/2}$ takes into account the finite mass effect.

\subsection{Scalar decay and absorption}

The Compton absorption $e+\phi \rightarrow e + \gamma$ process leads to the energy deposition inside the Borexino 
detector. Since only the sum of the deposited energy is measured, we would need a total cross section for this process. 
 The differential cross section we derive is the same as Eq. (5) of \cite{Izaguirre:2014cza} in the $m_\phi \ll E_\phi$ limit. 
 But in this paper we do not take the limit and use the full cross-section  $\sigma(e+\phi \rightarrow e + \gamma)$.
 The absorption length is then given by $L_{\rm abs}=1/(n_e \sigma_{e \phi\rightarrow e \gamma})$, where $n_e$ is the number density of 
 electrons inside the Borexino detector. It is easy to see that for the fiducial choice of parameters, the absorption length is much larger than the linear size of the detector. 
The Compton absorption process dominates in the very low $m_\phi$ regime, but the diphoton decays dominate in the medium and high mass range between a few hundred keV to 1.022 MeV (below pair production regime) as discussed below. 

The diphoton decay rate of the light scalar $\phi$ can be derived recasting the Higgs result \cite{Djouadi:2005gi},
\bea
\label{decay}
\Gamma (\phi \rightarrow \gamma\gamma)=
\frac{\alpha^2}{256\pi^3} m_\phi^3\left(\sum_{l=e,\:\mu,\:\tau} \frac{g_{\phi ll}}{m_l} \frac{2}{x_l^2}\biggl[x_l+(x_l-1)\arcsin^2(\sqrt{x_l})\biggr]\theta(1-x_l))\right)^2.
\eea
where $x_l=\frac{m_\phi^2}{4m_l^2}$ and $\theta(x)$ is the Heaviside step function. 
In principle, all charged particles with couplings to $\phi$ will contribute to the rate. Here we take into account only the charged leptons, while the inclusion of quarks would require additional information, beyond assuming a $g_p$ value. 
Therefore, this is an underestimation, 
with an actual $\Gamma (\phi \rightarrow \gamma\gamma)$ being on the same order but larger 
than (\ref{decay}). (One would need a proper UV-complete theory to make a more accurate prediction for the $\phi\to \gamma\gamma$ rate.)


When the mass of the scalar $m_\phi$ is larger than 2$m_e$, the electron-positron decay will dominate the diphoton and Compton absorption processes. We have
\begin{align}
\Gamma(\phi\rightarrow e^{+} e^{-})= \frac{g_e^2 m_\phi}{8\pi}\left(1-\frac{4m_e^2}{m^2_\phi}\right)^{3/2}.
\end{align}
The sum of these two rates determines the decay length, 
\begin{align}
{ L_{\rm dec}}=\beta\gamma\:(\Gamma(\phi\rightarrow e^{+} e^{-}) +\Gamma (\phi \rightarrow \gamma\gamma))^{-1},
\end{align}
where $\beta$ is the velocity of the scalar, which depends on its mass and energy, $\beta = \sqrt{1-m_\phi^2/E^2}$ ($c=1$ in our notations). 
The combination of absorption and decay,  
\bea
L_{\rm dec,\:abs} = (L_{\rm dec}^{-1}+ L_{\rm abs}^{-1}  )^{-1},
\eea
is required for the total event rate.

The decay/absorption length together with the geometric acceptance determines the probability of energy deposition inside 
the detector per each emitted scalar particle, 
\bea\label{P_dep}
P_{\rm deposit}&=&\int \frac{d(\theta)}{L_{\rm dec,\:abs}}\frac{2\pi}{4\pi} d \cos\theta\nn\\
&=& \frac{1}{L_{\rm dec,\:abs}}  \int^1_{\sqrt{1-(R/L)^2}} \sqrt{R^2-L^2(1-\cos^2\theta)}d\cos\theta\nn\\
&=& \frac{1}{L_{\rm dec,\:abs}} \times \frac{2 L R + (L^2 - R^2) \log\left( \frac{2 L}{L + R}-1\right)}{4 L },
\eea
where a spherical geometry of the detector is considered. 
Here $R$ is the fiducial radius and $L$ is the distance of the radiative source from the center of the detector. 
For our numerical results we use $R= 3.02$ m and $L= 8.25$ m as proposed in the SOX project \cite{Bellini:2013uui}.
In the $L\gg R$ limit, the probability has a simple scaling with the total volume and the effective flux at the position of the detector, 
\bea P_{\rm deposit} \simeq \frac{1}{L_{\rm dec,\:abs}}\frac{\frac{4}{3}\pi R^3}{4\pi L^2},\eea
but we use the complete expression (\ref{P_dep}) for the calculations below.

\subsection{Total event rate and sensitivity reach}\label{sec:reach_background}

Using formulae from the previous subsections, we can predict the signal strength as a function of $m_\phi$ and
coupling constants. The excited state of $^{144} {\rm Nd}$ has two gamma transitions,  $E_0=2.185$ MeV and $E_1=1.485$ MeV, partitioned with
${\rm Br}_0= 0.7$  and ${\rm Br}_1= 0.3$ branching ratios. Thus, the signal would constitute peaks at 2.185 MeV and 1.485 MeV in the Borexino spectrum. We assume that the normal gamma quanta of these energies are efficiently degraded/absorbed by shielding. 

The signal counting rate for  a light scalar $\phi$ of energy $E_i$ (2.185 MeV or 1.485 MeV as 
$i=0$ or 1) in the Borexino detector is given by
\bea
\dot{N}_{ i}&=&\left(\frac{d N}{dt}\right)_{0}\exp\left(-\frac{t}{\tau}\right) \times {\rm Br}_{^{144}{\rm Nd}^*} \times 
{\rm Br}_i \times
\frac{1}{2}
 \left(\frac{g_p}{e}\right)^2 \left(1-\left(\frac{m_\phi}{E_i}\right)^2\right)^{3/2}
\times
P_{{\rm deposit},\:i}.
\label{count}
\eea
Here, $\left(\frac{d N}{dt}\right)_{0}$ is the initial source radioactivity in units of decays per time, and the projected strength is
$\simeq$ 5 PBq, or $5\times 10^{15}$ decays per second. $\tau$ is the lifetime of $^{144}$Cr, $\tau = 285$ days. ${\rm Br}_{^{144}{\rm Nd}^*} $ is the probability that the 
$\beta$-decay chain 
leads to the 2.185 MeV excited state of $^{144}{\rm Nd}$, ${\rm Br}_{^{144}{\rm Nd}^*} \simeq 0.01$. Finally, $P_{{\rm deposit},\:i}$ is the probability of decay/absorption defined in the previous subsection 
that depends on $i$ via the dependence of the decay length and the absorption 
rate on $E_i$. Substituting relevant numbers we get the counting rate for the 2.185 MeV energy as 

\bea
\dot{N}_{\rm 2.185\;\rm MeV}\left[\frac{\rm counts}{\rm day}\right]&=&1.5\times 10^{18}\times \exp\left(-\frac{t[\rm day]}{285d}\right)
\times \frac{\left(\frac{d N}{dt}\right)_{\rm 0}}{5 \rm PBq} 
\nn
\\
&&
\times
 \left(\frac{g_p}{e}\right)^2 \left(1-\left(\frac{m_\phi}{2.185 \rm MeV}\right)^2\right) ^{3/2}
\times
P_{\rm deposit,\:\rm 2.185\:\rm MeV} 
\eea

The resulting sensitivity reach of the three processes considered is plotted in the left panel of Fig. \ref{fig:sen_con} as a blue curve.
Here we assume the mass-proportional coupling strengths for $\phi$ to proton and leptons, and parametrize the coupling as 
$\epsilon^2=g_p g_e/ e^2.$ The curve corresponds to a $3\:\sigma$ sensitivity level with the assumption that the initial 
source strength is 5 PBq.

For the derivation of the future sensitivity reach, we have followed the simplified procedure:
For every point on the parameter space $\{m_\phi,~\epsilon\}$, we calculate the expected counting rate using Eq. (\ref{count}).
We then take an overall exposure of $t_{exp}=365$ days to arrive at an expected number of signal events as a function of mass and coupling, $N_{sig}(m_\phi,\epsilon)$. The background is the total number of events in energy bins near $E=2.185$ MeV and $E=1.485$ MeV.
The energy resolution at Borexino is $5\%\times \sqrt{1{\rm MeV}/E}.$ We use this as the bin size when we estimate the background rates at $E=2.185$ MeV and $E=1.485$ MeV.
For the background event rate, we use the energy spectra shown in Fig. 2 in \cite{Bellini:2013uui}. After all cuts, the background rate is $R_{backgr} \simeq$ 200 counts/100t$\times$100keV per 446.2 live-days at energy $E=2.185$ MeV (For $E=1.485$ MeV, the background rate is around 2300 counts/100t$\times$100keV). For $E=2.185$ MeV this gives the total number of background events to be $N_{backgr} \simeq 90$.
We then require $N_{sig} < 3\sqrt{N_{backgr}}$ that results in the sensitivity curve in Fig. 1. Based on our estimation, the inclusion of $E=1.485$ MeV channel does not lead to a significant improvement: it allows one to increase the significance by roughly $0.2\sigma$ with respect to just considering the main 2.185 MeV channel. Should a strong signal be observed, however, the presence of two peaks would be an unmistakable signature. 


In the above procedure, we have taken into account only the existing source-unrelated backgrounds.
However, a question arises whether additional inverse beta decay (IBD) events in Borexino, $p+\bar \nu \to n +e^+$, which is the primary goal of the SOX project, may also affect the search for $E=2.185$ MeV abnormal energy deposition.
If the location of IBD event is inside the fiducial volume, then even the threshold IBD event creates 3.2 MeV energy deposition.
(The positron at rest produces 1.0 MeV energy, and the neutron capture results in the additional 2.2 MeV).
This is well outside the energy windows for the signal from exotic scalars. Moreover, IBD events have a double structure in time, which can be used to discriminate them.
An interesting question arises whether the location of IBD events outside the fiducial volume ({\em i.e.} close to the edge of the detector) may lead to a loss of positron signal followed by the neutron capture inside the fiducial volume. For a neutron with a typical kinetic energy of a keV would have to diffuse for at least 1m inside liquid scintillator to reach the fiducial volume. However, the estimates of Ref. \cite{Vogel:1999zy} show that the typical diffusion length is $O(5\,{\rm cm})$, which render the probability for such events to be small. 

%
Still, background events could occur when the neutron-proton capture takes place in the non-scintillating buffer region at a radius $ R > 4.25 \rm\: m$, if the 2.2 MeV capture gamma ray (with attenuation length $\sim$ 90 cm) reach the fiducial volume  at $ R < 3 \rm\:m$ and mock the 2.185 MeV signal. 
In regard of this potential background, we conduct an additional analysis taking a fiducial radius ($R = 2.00 \rm\: m$) smaller than the $R = 3.02 \rm\: m$ used in the Borexino analysis \cite{Bellini:2013uui}. We plot both the sensitivity reaches based on 2.00 m and 3.02 m fiducial radii in Figure 1 and 2. One can regard the sensitivity reach with 2.00 m fiducial radius a more conservative estimation.
Furthermore, this gamma-ray background would have to appear in a radial dependent fashion in the detector, meaning that the background is stronger in the regime nearer to the buffer area.
Such information on radius dependence can be applied to further subtract the background events. We leave the simulation to accurately determine this background to future works.


To be more inclusive, we also consider a variant of the scalar model when the couplings to electrons and tauons are switched off (muonic scalar). In this case, the remaining energy depositing channel is the diphoton decay, and there is no gain in sensitivity for $m_\phi > 2 m_e$. We plot the corresponding sensitivity reach in the right panel of Fig. \ref{fig:sen_con} also as a blue curve.

\section{Comparing to existing constraints}
\label{sec:cons}

Here we reassess some limits on the couplings of very light scalars. The most significant ones are from the beam dump experiments, meson decays and stellar energy losses. The particle physics constraints that rely on flavor changing processes are difficult to assess, as they would necessarily involve couplings of $\phi$ to the heavy quarks. We leave them out as model-dependent constraints.

\subsection{Beam dump constraints}

Among the beam dump experiments, the LSND is the leader given the number of particles it has put on target. 
The LSND measurements of the elastic electron-neutrino cross section 
\cite{Auerbach:2001wg, Aguilar:2001ty} can be recast to put current-leading constraints on the parameter spaces of our model, as well as models including light dark matter and millicharged particles \cite{Kahn:2014sra,Magill:2018tbb}, and models with neutrino-heavy neutral lepton-photon dipole interactions \cite{Magill:2018jla}.
Here we revise previous bounds discussing different production channels, and account for scalar decays and Compton absorptions inside the LSND volume. 

The collisions of primary protons with a target at LSND energies produce mostly pions and electromagnetic radiations. 
Exotic particles, such as scalars $\phi$ can be produced in the primary proton-nucleus collisions, as well as in the subsequent decays and 
absorptions of pions. A detailed calculation of such processes would require a dedicated effort. It would also require more knowledge 
about an actual model, beyond the naive Lagrangian (\ref{lag_rp}). In particular, one would need to know how the scalars couple to 
pions and $\Delta$-resonances, that alongside nucleons are the most important players in the inelastic processes in the LSND experiment 
energy range. 
Here we resort to simple order-of-magnitude estimates, assuming that the $g_p$ coupling is the largest, and drives the production
of scalars $\phi$.
 
The important process for the pion production at LSND is the excitation of $\Delta$ resonance in the collisions of 
incoming protons with nucleons inside the target. Assuming that the decay of $\Delta$'s saturates the pion 
production inside the target, we can estimate the associated production of scalars in the $\Delta \to p +\pi +\phi$ process. 
To that effect, we consider the following two interaction terms, 
\begin{equation}
\label{Delta}
{\cal L}_{int} \sim g_p \phi \bar p p +  g_{\pi\Delta p}(\Delta_\mu p )\partial_\mu  \pi,
\end{equation}
where $\Delta_\mu$ is the Rarita-Schwinger spinor of $\Delta$-resonance, $g_{\pi\Delta p}$ is the pion-delta-nucleon 
coupling constant, and the isospin structure is suppressed. To estimate scalar production, we calculate the 
rates for $\Delta \to p +\pi $, $\Delta \to p +\pi +\phi$ and take the ratio finding
\begin{equation}
\label{yield_De}
N_\phi \sim N_\pi \times \frac{\Gamma_{\Delta \to p \pi\phi}}{\Gamma_{\Delta \to p \pi} }\simeq N_\pi \times 0.04 g_p^2.
\end{equation}
Notice that the decay rates are relatively large, being enhanced by the $\log(Q/m_\phi)$, where $Q$ is the energy release. 
The coefficient 0.04 is calculated for $m_\phi=1$ MeV, and it varies from 0.06 for $m_\phi = 0.2$ MeV to 0.03 $m_\phi = 2$ MeV. 


Depending on their charges, pions have very different histories inside the target. 
The negatively charged  $\pi^-$ undergoes nuclear capture. In \cite{Izaguirre:2014cza} the rate of the scalar 
production in nuclear capture was overestimated, as it was linked to the production of photons in the capture of $\pi^-$ by 
free protons via $e^2 \to g_p^2$ substitution. 
The radiative capture rate on protons is about 40\%. 
For the LSND target, however, the more relevant process is the radiative 
capture on nuclei with $A\geq 16$, which is in the range of $\sim 2\%$ \cite{Amaro:1997ed}. Therefore, 
one may use 
$N_\phi(\pi^-) \sim 0.02 \times N_{\pi^-} \times \left( \frac{g_\pi}{e} \right) ^2$ as an estimate for the production rate of scalars from the $\pi^-$ capture. Notice this is the coupling of scalars to pions that mostly determines the capture rate. Moreover, the number of $\pi^-$ 
is smaller than the total pion production,  and therefore we expect the production of $\phi$ in the  $\pi^-$ capture to be subdominant 
to $\Delta$ decays (\ref{yield_De}). 
Unlike the case with negatively charged pions, most of $\pi^+$ stop in the target and decay. The scalar particle is then produced in the three-body decay, $\pi^+\to \mu^+\nu \phi$, and in the four-body decay of the stopped $\mu^+$, $\mu^+ \to e^+\nu\nu\phi$. The decays of $\pi^0$ are instantaneous, and they could also lead to the production of light scalars in $\pi^0\to \gamma\gamma\phi$. Direct estimates of the corresponding branching ratios give $\sim 0.05 (g_{\mu(\pi)})^2$, and again we find that this is subdominant to (\ref{yield_De}) estimate because of $g_{\mu(\pi)}< g_p$. 


A conservative estimate of the number of pions produced in the experiment is $N_{\pi}\sim 10^{22} $ 
(see, {\em e.g.}, \cite{Auerbach:2001wg}). We take 300 MeV as an estimate for the average energy of scalars.
Now we can estimate the expected number of events $N_{\rm LSND}$, {\em i.e.}
the number of light scalars that deposit their energies in the LSND detector:
\begin{align}
\nonumber N_{\rm LSND}\sim & 
N_{\pi} \times 0.04\:g_p^2 \times
  P_{\rm\;survive\;+\;deposit\;in\;LSND}\\
  \simeq&
  N_{\pi} \times 0.04\:g_p^2 \times
\left[\exp\left(-\frac{L_{\rm LSND}-\frac{d_{\rm LSND}}{2}}{L_{\rm dec}}\right)-\exp\left(-\frac{L_{\rm LSND}+\frac{d_{\rm LSND}}{2}}{L_{\rm dec,\; abs}}\right)\right]
\left(\frac{A_{\rm LSND}}{4\pi L^2_{\rm LSND}}\right).
\end{align}
Here we conservatively assume spatially isotropic distribution, take $L_{\rm LSND}=30$ m as the distance between the target and the center of the detector, $d_{\rm LSND}=8.3$ m is the length of the detector itself, and $A_{\rm LSND}\simeq 25\rm\:m^2$ is the cross-section of the detector looking from the side \cite{Athanassopoulos:1996ds, Auerbach:2001wg}. $ L_{\rm dec, abs}$ are the decaying and absorption length determined by the physical processes Eq. (\ref{three_processes}). 
Notice that we no longer use the assumption that $L_{\rm dec,\; abs}\gg L_{\rm LSND}{\rm \;and\;}d_{\rm LSND}$ since in the high $\epsilon^2$ regime these three lengths could be comparable. The number density of electrons in the LSND detector is $n_e=2.9 \times 10^{29}$ $\rm m^{-3}$, and the absorption again plays a subdominant role in the energy deposition process.



\begin{figure}[t]
\begin{center}
\includegraphics[width =0.49
 \textwidth]{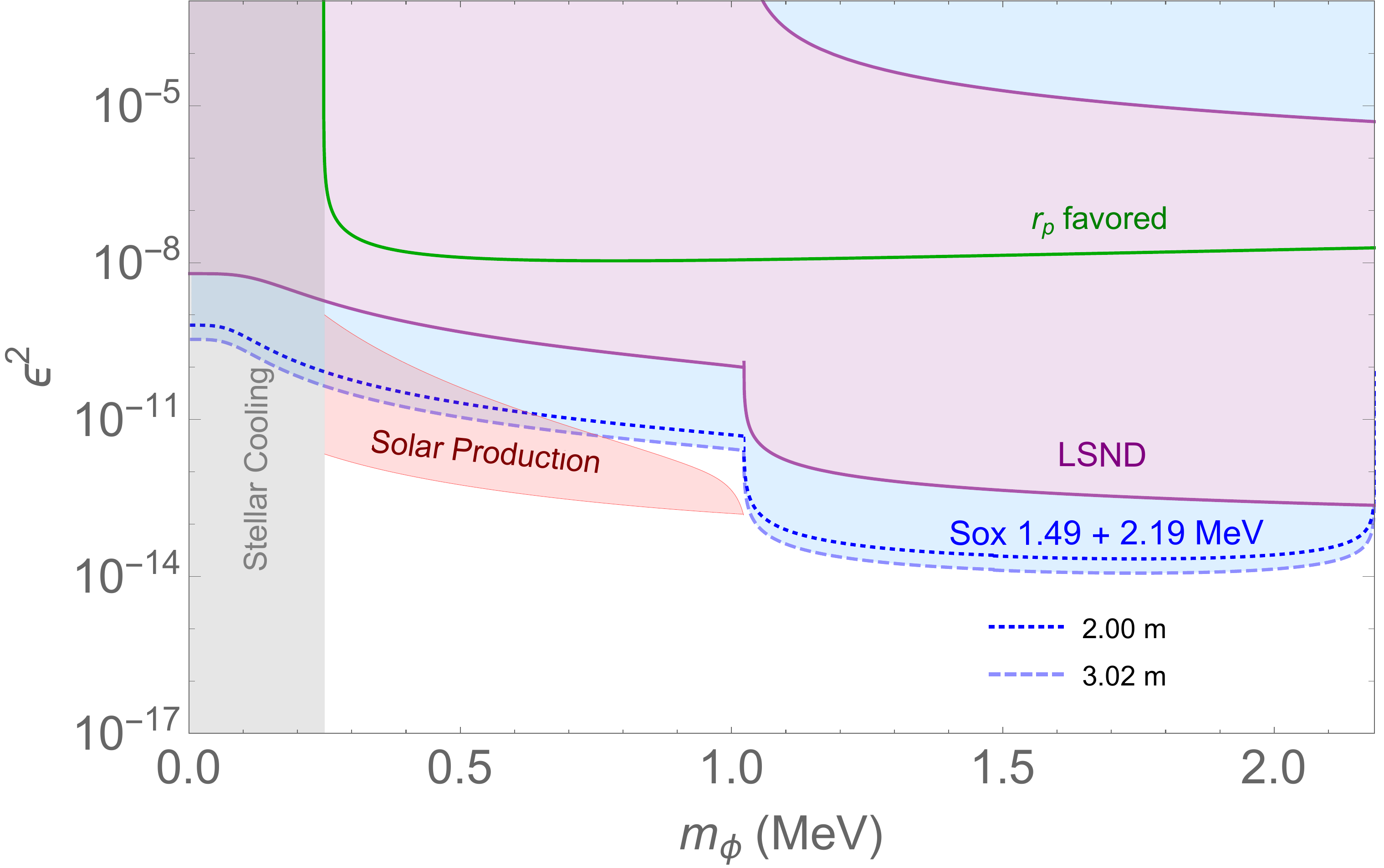}
\includegraphics[width =0.49
 \textwidth]{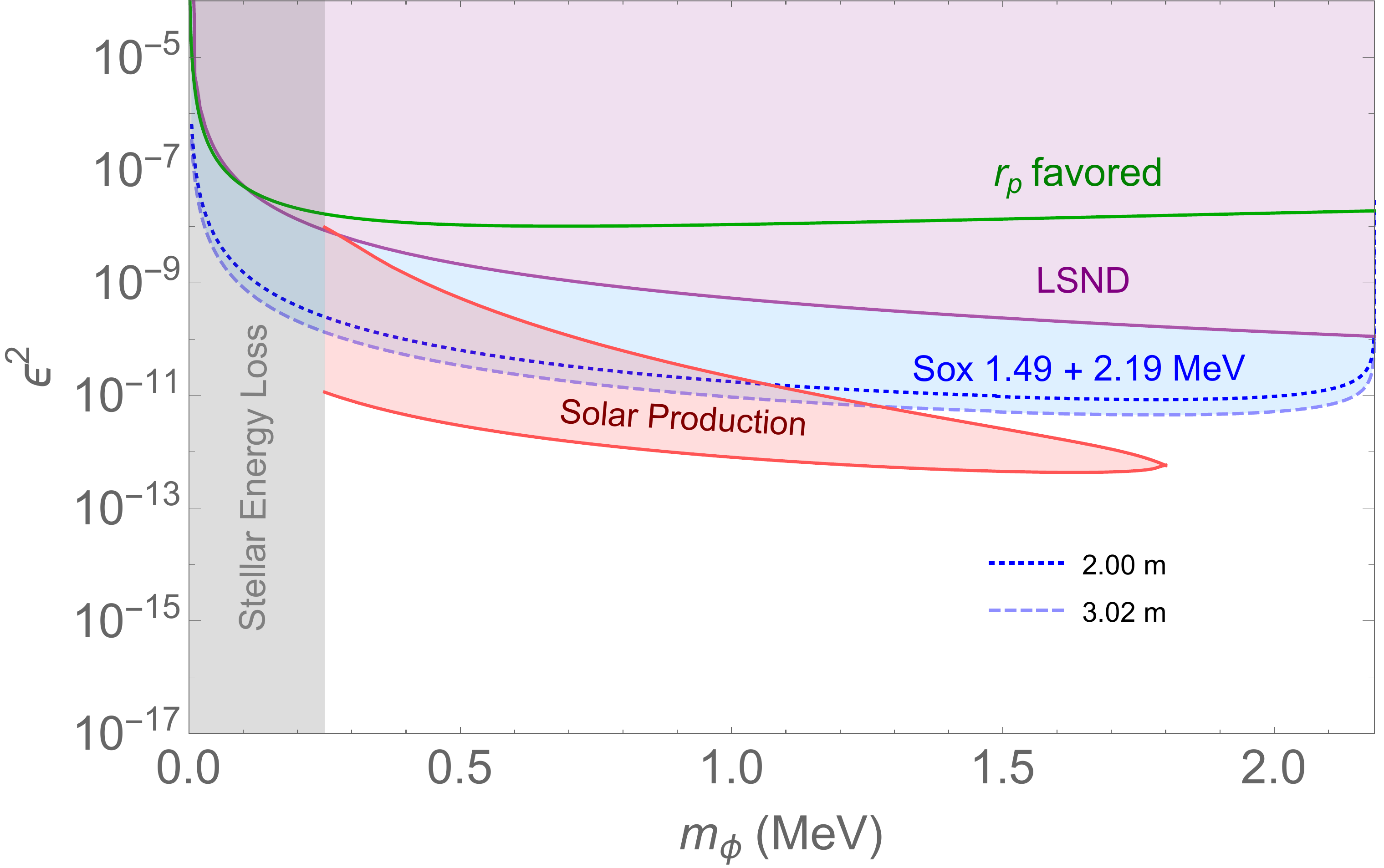}
\caption{ 
Future sensitivity reach of the Borexino-SOX setup and existing constraints placed on the coupling constant-mass parameter space. 
We conduct the analysis in two fiducial radii, 2.00 m and 3.02 m, for the Borexino-SOX sensitivity reaches, in regard of the background from the 2.2 MeV n-p capture gamma ray discussed in section \ref{sec:reach_background}. 
{\em Left panel:}
The $g_i \propto m_i$ scaling is assumed and $\epsilon$ is defined as $\epsilon^2 = g_pg_e/e^2$.
{\em Right panel:} $g_e=g_\tau=0$, a $g_i \propto m_i$ scaling for $\mu$ and $p$, while $\epsilon^2 = (m_e/m_\mu)\times g_pg_\mu/e^2$. The green curve is the parameter space that can explain the proton-size anomaly.
The experimental reach ($> 3\sigma$) by the Borexino-SOX setup is the blue regime.
The recast of LSND constraints \cite{Auerbach:2001wg} is shown in purple, while the gray area is constrained by the stellar energy loss \cite{Raffelt:1994ry}.
The solar production constraint \cite{Smirnov:2015lxy} is the protruding pink area between $\epsilon^2=10^{-9}$ and $10^{-13}.$}
\label{fig:sen_con}
\end{center}
\end{figure}

Based on Fig. 10 of \cite{Auerbach:2001wg} and Fig. 28 of \cite{Aguilar:2001ty} we estimate that there are less than 20 decay-in-flight
 events above 140 MeV during the exposure. 
 We then determine the LSND constraint on the parameter space of the $\phi$ scalar as plotted in Fig. \ref{fig:sen_con} in purple color.
We reiterate a rather approximate nature of the estimates. 

\subsection{Solar emission and stellar energy loss}

Thermal production of scalars may lead to abnormal energy losses (or abnormal thermal conductivity) 
that would alter the time evolution of well known stellar populations. 
In the regime of $m_\phi >T$, the thermally averaged energy loss is proportional to $g_e^2\:{\rm exp}(-m_\phi/T_{\rm star}).$ 
Given the extreme strength of stellar constraints  \cite{Raffelt:1994ry}, one can safely exclude $m_\phi < 250 $ keV  
for the whole range of coupling constants considered in this paper. 

In addition, the non-thermal emission of scalars in nuclear reaction rates in the Sun can also be constrained. 
The light scalar $\phi$ can be produced in the Sun through the nuclear interaction $p+$D$\rightarrow {\rm^3 He} + \phi.$ This process generates a 5.5 MeV $\phi$ flux that was constrained by the search conducted by the Borexino experiment. The flux can be estimated as 
\begin{align}
\Phi_{\phi, \rm solar} \simeq (g_p/e)^2 \Phi_{pp\nu} P_{\rm esc} P_{\rm surv}. 
\end{align}
Here $\Phi_{pp\nu}= 6.0 \times 10^{10}$ cm$^{-2} s^{-1}$ is the proton-proton neutrino flux. $P_{\rm esc}$ is the probability of the light scalar escaping the Sun while $P_{\rm surv}$ is the probability of the scalar particle not decay before it reaches the Borexino detector.
\begin{align}
P_{\rm esc}&=\exp\left(-\int^{R_{\small \astrosun}} dr\: n_{ \small \astrosun} \sigma_{e\phi\rightarrow e\gamma} \right)\\
P_{\rm surv} &= \exp\left(- \frac{L_{\small \astrosun}}{L_{\rm dec}}\right)
\end{align}
where $R_{\small \astrosun}$ and $L_{\small \astrosun}$ are the radius of the Sun and Earth-Sun distance respectively, while
$n_{\small \astrosun}$ is the mean-solar electron density.

$L_{\rm dec}$ is again determined by the decay processes in Eq. \ref{three_processes}. For $m_\phi<2m_e$ the $\phi$ particle can survive and reach the Borexino detector when $\epsilon<10^{-9}$, and deposit its energy through processes in Eq. (\ref{three_processes}).
For $m_\phi>2m_e$ the $P_{\rm surv}$ is highly suppressed due to rapid di-electron decays and thus $m_\phi=2m_e$ is where the constraint ends. 

Notice that it is difficult to impose the supernovae (SN) constraints on this model, because of the uncertainties in the choices of some couplings. In general, we believe that the coupling of scalars to nucleons can be large enough so that they remain trapped in the explosion zone, therefore avoiding the SN constraint. 

\section{Sensitivity to dark photons below 1 MeV} 
\label{sec:dph}

Dark photon is a massive ``copy'' of the regular SM photon, which couples to the electromagnetic current with a strength proportional to a small mixing angle $\epsilon$, realized as a kinetic mixing operator. The low-energy Lagrangian for dark photons can be written as
\begin{equation}
{\cal L}_{\rm d.ph.} = -\frac{1}{4} F'_{\mu\nu} F'^{\mu\nu} +\frac{1}{2}m^2_{A'}(A'_\mu)^2 + \epsilon A'^\mu J_\mu^{EM}.
\label{dph}
\end{equation}
Here $J_\mu^{EM}$ is an operator of the electromagnetic current. 

This model is very well studied, and in many ways, it is more attractive than the model of scalars in (\ref{lag_rp}) mainly because it has a natural UV completion. Zooming in on the parameter space relevant for the Borexino-SOX, we discover that above $2 m_e$ the combination of all beam dump constraints put strong limits on the dark photon model. For $m_{A'} < 2 m_e$ the most challenging constraint comes from cosmology, where the inclusion of three $A'$ polarizations, fully thermalized with electron-photon fluid, will reduce the effective number of neutrino species to an unacceptable level $N_{eff} <2$ \cite{Nollett:2013pwa}. Only a judicious choice of additional ``passive'' radiation could put this model back into agreement with cosmology.

Fully realizing all the complications coming from cosmology, we nevertheless estimate the sensitivity of the proposed setup to $\epsilon$.
An interesting feature of the dark photon model below the $2m_e$ threshold is that the main decay channel is $3\gamma$, and it is mediated by the electron loop. The decay rate is very suppressed, and the effective-field-theory type calculation performed in the limit of very light $A'$ \cite{Pospelov:2008jk} was recently generalized to the $m_{A'} \sim 2 m_e$ \cite{McDermott:2017qcg}. We take this decay rate, and in addition, calculate separately the cross section of the scattering process $e +A' \to e + \gamma$.
Due to the strong suppression of the loop-induced decay, we find that the Compton-type scattering gives the main contribution to the signal rate in Borexino.

For the dark photon $A'$, the emission rate 
(the rate of the nuclear state decay to $A'$) is determined by 
\begin{equation}
\frac{\Gamma_{A' }}{\Gamma_{\gamma,~E1} } = \frac{v_{A'}(3-v_{A'}^2)}{2} \epsilon^2,
\end{equation}
where $v_{A'}= (1-m_{A'}^2/\omega^2)^{1/2}$. In the limit of $m_{A'}\ll \omega$, the ratio of the two rates becomes simply $\epsilon^2$. 
Substituting relevant numbers we get the counting rate for the 2.185 MeV energy as 

\bea
\dot{N}_{A',\;\rm 2.185\;\rm MeV}\left[\frac{\rm counts}{\rm day}\right]&=&1.5\times 10^{18}\times \exp\left(-\frac{t[\rm day]}{285d}\right)
\times \frac{\left(\frac{d N}{dt}\right)_{\rm 0}}{5 \rm PBq} 
\nn
\\
&&
\times
\epsilon^2\times
 v_{A'} (3-v_{A'}^2) 
\times
P_{\rm deposit,\:\rm 2.185\:\rm MeV} 
\eea

For the background event rate, we use the energy spectra shown in Fig. 2 in 
\cite{Bellini:2013uui}. We use the 4th/green event spectrum with the fiducial volume (FV) cut. The background rate is around 200 counts/100t$\times$100keV per 446.2 live-days at energy $E=2.185$ MeV (For $E=1.485$ MeV, the background rate is about 2300 counts/100t$\times$100keV).
We got the sensitivity curve in Fig. 2 by having an overal exposure of $t_{exp}=365$ days and consider the coupling $\epsilon$ for each mass that gives $N_{sig} < 3\sqrt{N_{backgr}}$.
The energy resolution at Borexino is $5\%\times \sqrt{1{\rm MeV}/E}.$ We use this as the bin size when we estimate the background rates at $E=2.185$ MeV and $E=1.485$ MeV.
In regard of the background from the 2.2 MeV n-p capture gamma ray discussed in section \ref{sec:reach_background}, we again conduct the analysis in two fiducial radii, 2.00 m and 3.02 m, for the Borexino-SOX sensitivity reaches.

Even though the particle $A'$ cannot decay to $e^+e^-$ in the kinematic range we consider, the decays to photons and the Compton-like absorption will lead to the beam dump constraints for this model. The LSND production is easy to estimate, 
given that $\pi^0$ will always have an $A'\gamma$ decay mode with $Br_{\pi^0 \to A'\gamma} = 2 \epsilon^2$. 

\begin{figure}[t]
\begin{center}
\includegraphics[width =0.70
 \textwidth]{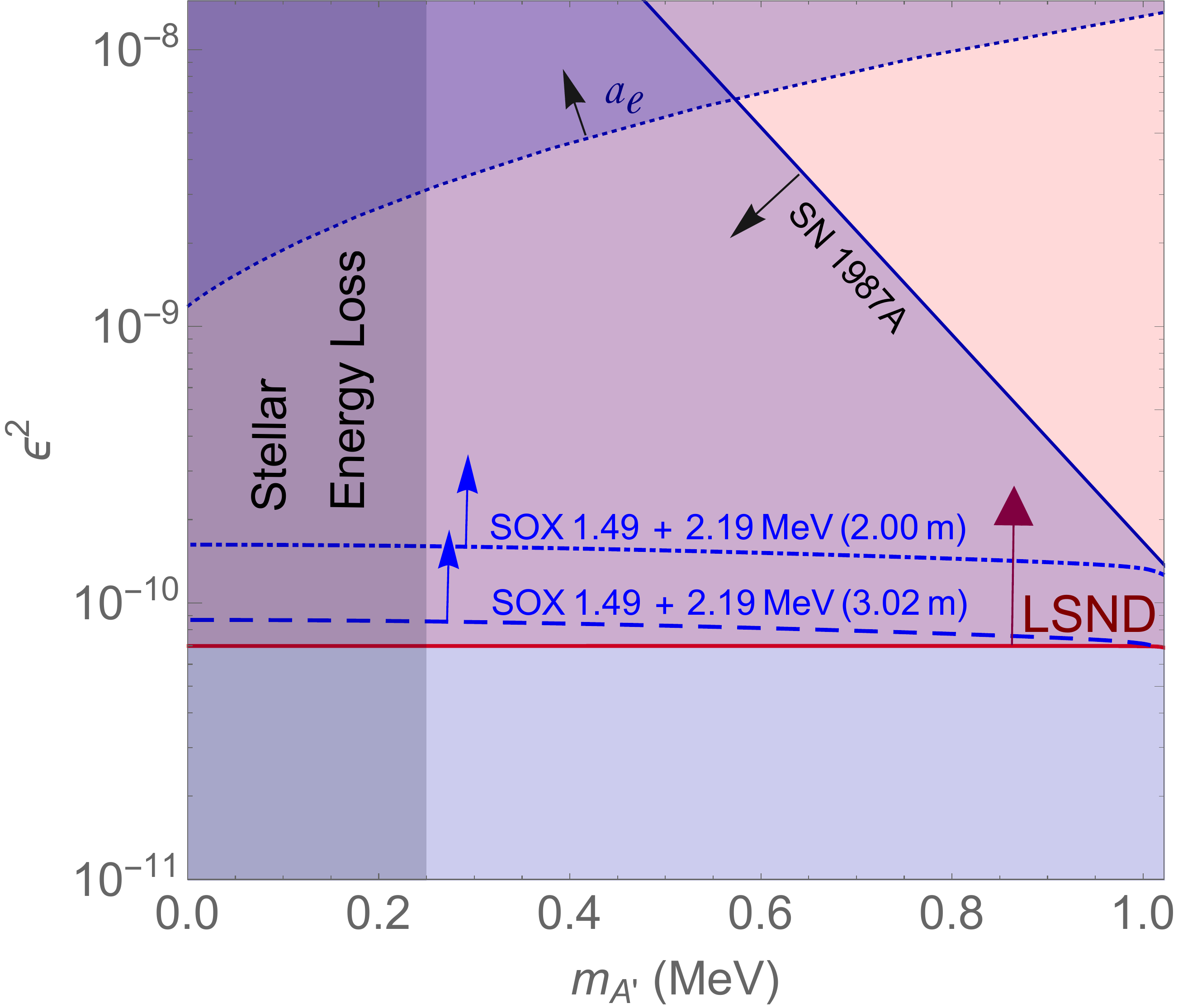}
\caption{ 
{
Future sensitivity reach for the Borexino-SOX setup and various existing constraints in coupling constant-mass parameter space for dark photons with a small mixing angle $\epsilon$.
Again, we conduct the analysis in two fiducial radii, 2.00 m and 3.02 m, for the Borexino-SOX sensitivity reaches, in regard of the background from the 2.2 MeV n-p capture gamma ray discussed in section \ref{sec:reach_background}.
{\em Left panel:}
The experimental reach ($> 3\sigma$) by the Borexino-SOX setup is the blue curve. The constraint recasting the LSND data \cite{Auerbach:2001wg} is slightly stronger than the Borexino-Sox reach, and excludes all the parameter space above the purple curve.
Supernova cooling constrains the whole regime below the dark blue curve on the upper-right corner \cite{Chang:2016ntp, Hardy:2016kme}, while the gray area is again the stellar energy loss bound \cite{Raffelt:1994ry}.
}
}
\label{fig:sen_con_dph}
\end{center}
\end{figure}

A compilation of all the considerations above is shown in Fig. \ref{fig:sen_con_dph}. 
We find that the sensitivity reach of the Borexino-SOX experiment, $\epsilon^2 \sim 10^{-10}$ in probing the light dark photons, is comparable but slightly above the bound from recasting the LSND data. Furthermore, this LSND bound covers up a small triangular parameter space for $10^{-5}\le\epsilon\le10^{-4},\; m_A'\le 2m_e$ that was not excluded by the cooling of Supernova 1987A \cite{Chang:2016ntp, Hardy:2016kme}, and the precision measurement of electron anomalous magnetic moment (see Fig. 7 of \cite{Chang:2016ntp}), independently from the cosmological scenarios. 
Note that here we plot the ``robust'' constraint from \cite{Chang:2016ntp} in our Fig. \ref{fig:sen_con_dph}, which is the intersection of bounds from different supernova profile models. 
Also, both \cite{Chang:2016ntp, Hardy:2016kme} use the trapping criterion, rather than the energy transport criterion (see, {\em e.g.}, \cite{Burrows:1990pk,Keil:1994sm}), to set the upper limits for the SN exclusion regions, with the trapping criterion being more conservative.

\section{Conclusion}
\label{sec:conclusion}

We have considered in detail how the search of the sterile neutrinos in the Borexino-SOX experiment can also be turned into a search for extremely weakly interacting bosons. The reach of the experiment to the parameters of exotic scalars is limited by the energy release in radioactive cascades. It has to be less than 2.185 MeV for the radioactive source to be used in SOX. However, in terms of the coupling constants, the reach of this experiment will be much farther down than even the most sensitive among the particle beam dump experiments. 
We find that with the proposed setup, coupling constants as low as $\epsilon^2 \sim 10^{-14}$ will be probed.
The improved analysis in this work includes particle decays inside the detector as the main energy-deposition channel.
It is the dominant process that significantly exceeds the scalar Compton absorption above the hundred-keV mass regime. Similar revisions will apply to searches proposed in Ref. \cite{Izaguirre:2014cza} that suggest using proton accelerators to populate nuclear metastable states.
In addition, we study the sensitivity reach of the Borexino-SOX experiment in probing a light dark photon below 1 MeV. The reach $\epsilon^2 \sim 10^{-10}$ is comparable, but slightly weaker than the bound already imposed by the existing LSND neutrino-electron scattering data.
Combining this constraint with the supernova bound we completely rule out the possibility of having a light dark photon below 1 MeV in this coupling range. 

In conclusion, one should not regard the SOX project as exclusively a search for sterile neutrinos (motivated mostly by experimental anomalies), but a generic search for dark sector particles. The scalar case considered in this paper can be motivated by the proton charge radius anomaly, and the SOX project provides tremendous sensitivity to this type of models. We encourage the Borexino collaboration to perform its own study of the sensitivity to new bosons using more detailed information about background and efficiencies.

\acknowledgments 
We thank Drs. P. deNiverville and S. Zavatarelli for useful correspondence. We also thank Drs. J. Dror, R. Lasenby and B. Safdi for helpful discussions. Research at the Perimeter Institute is supported in part by the Government of Canada through NSERC and by the Province of Ontario through MEDT. YT was supported by the Visiting Graduate Fellow program at Perimeter Institute, U.S. National Science Foundation through grant PHY-1719877, and Cornell graduate fellowship, while parts of this work were completed.

\bibliography{borex_ref}{}

\end{document}